\documentclass[aps,prl,onecolumn,superscriptaddress,amsmath,amssymb,nofootinbib]{revtex4}
\usepackage{graphicx}
\usepackage{bm}
\allowdisplaybreaks

 \newcommand{\be}{\begin{equation}}
 \newcommand{\ee}{\end{equation}}
\newcommand{\bea}{\begin{eqnarray}}
\newcommand{\eea}{\end{eqnarray}}
\newcommand{\ba}{\begin{eqnarray*}}
\newcommand{\ea}{\end{eqnarray*}}

\newcommand{\br}{\mathbf{r}}

\usepackage{multirow}

\usepackage[version=3]{mhchem} 
\usepackage[english]{babel}
\usepackage{dcolumn}
\usepackage[colorlinks=true,linkcolor=blue,citecolor=blue,
bookmarksnumbered=true,bookmarks=true,plainpages=false,pagebackref]{
hyperref}

\begin{document}

\title[Distinct metallization and atomization transitions in dense liquid hydrogen]
{Distinct metallization and atomization transitions in dense liquid hydrogen}

\author{Guglielmo Mazzola}
\email{gmazzola@sissa.it}
 \affiliation{International School for Advanced Studies (SISSA), and CRS Democritos, CNR-INFM, - Via Bonomea 265, I-34136 Trieste, Italy}

\author{Sandro Sorella}
\email{sorella@sissa.it}
\affiliation{International School for Advanced Studies (SISSA), and CRS Democritos, CNR-INFM, - Via Bonomea 265, I-34136 Trieste, Italy}

\date{\today}
             
\begin{abstract}
We perform molecular dynamics simulations driven by accurate Quantum Monte Carlo forces on dense liquid hydrogen.
Recently it has been reported  a complete atomization transition between a mixed-atomic liquid and a completely dissociated fluid in an almost unaccessible pressure range \mbox{[\emph{Nat. Commun.} {\bf 5}, 3487 (2014)]}.
Here instead, in a  much more interesting pressure range, we identify a different transition between the fully molecular liquid and the mixed-atomic fluid at $\sim$ 400 GPa, with numerical evidence supporting its metallic behavior.
Therefore we predict that the metallization at finite temperature occurs in this partially dissociated molecular fluid, well before the complete atomization of the liquid.
At high temperature this first-order transition becomes a crossover, in very good agreement with the experimental observation.
Several systematic tests supporting the quality of our large scale calculations are also reported.

\end{abstract}
  \maketitle
  
\begin{figure}[htp]
 \includegraphics[width=1.0\columnwidth]{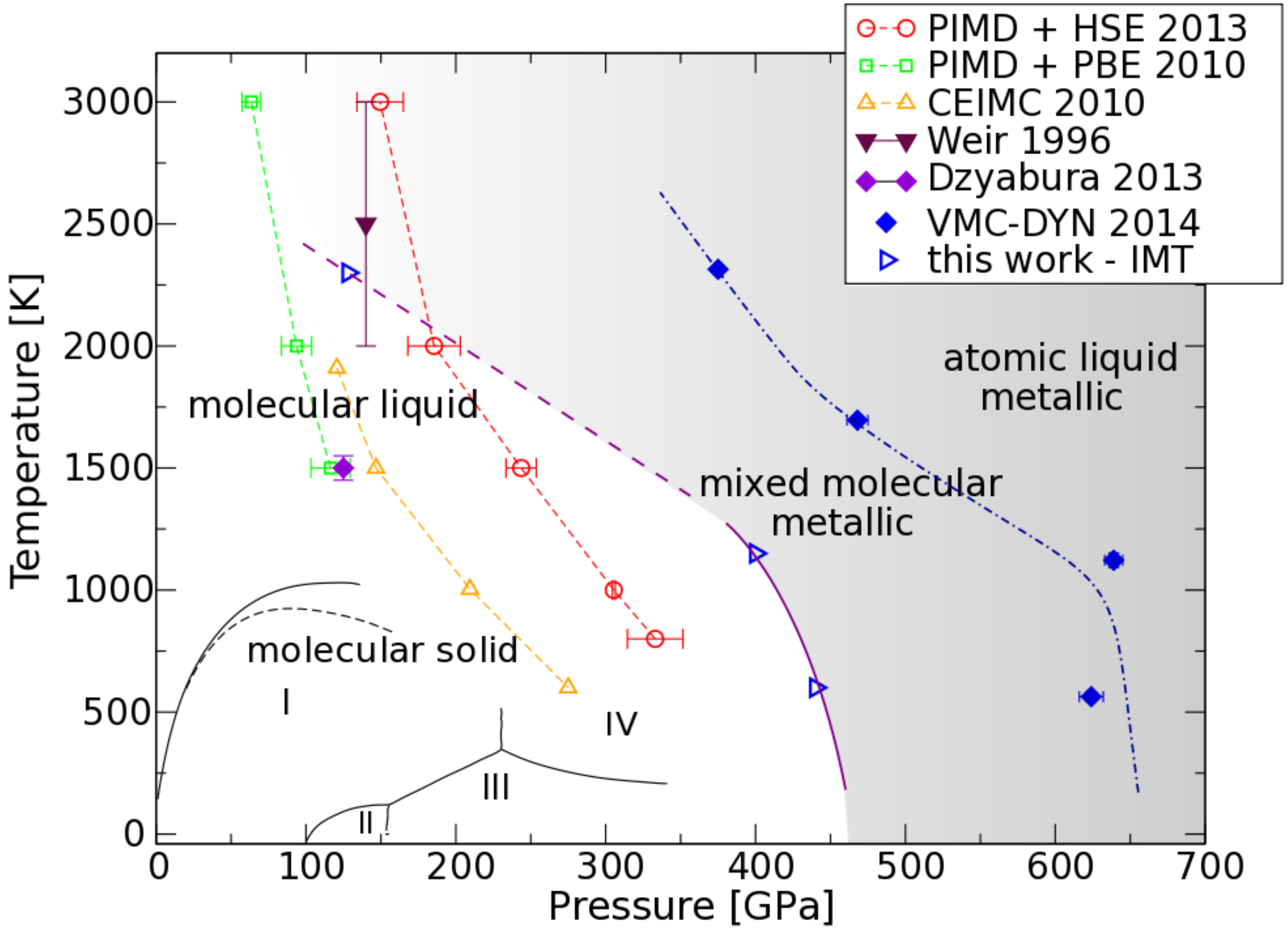}
 \caption{ (color online). $P$-$T$ phase diagram of hydrogen. Black solid lines indicate experimental phase boundaries between the molecular liquid and the different molecular solid phases (I, II, III, and IV) as in Ref.~\cite{zha_synchrotron_2012}. Colored symbols correspond to the liquid-liquid transition obtained with latest simulations and experimental points. Red circles and green squares refer to Density Functional Theory (DFT) calculations with different functionals (PBE and HSE) and including proton quantum effects~\cite{morales_nuclear_2013} while orange triangles refer to Coupled Electron-Ion Monte Carlo (CEIMC)~\cite{morales_evidence_2010,liberatore_free_2011}. Blue solid diamonds correspond to the liquid-liquid transition (LLT) estimated in Ref.~\onlinecite{mazzola_unexpectedly_2014} (VMC-DYN). The (blue dash-dotted) LLT lines divides the completely dissociated fluid (on the right) from the molecular/partially dissociated fluids (on the left). Blue triangles are instead placed at the observed transition in this work, obtained by looking at the density matrix and  the EOS discontinuities.  The shaded region corresponds to the metallic liquid. The IMT is first order below (above) $\sim$ 1100 K and is indicated with a solid (dashed) purple line, which is a guide to the eye.
 }
 \label{f:phase}
 \end{figure}

In spite of seemingly a simple system to study, hydrogen displays a very rich phase diagram that, in most cases, it is not completely determined from an experimental point of view\cite{mcmahon_properties_2012,howie_mixed_2012,zha_synchrotron_2012,zha_high-pressure_2013}.
At low pressures and temperatures, hydrogen crystallizes as a molecular solid and undergoes a number of phase transitions with increasing density. 
Up to the present days, this material has been probed experimentally, at low temperatures, up to the pressure of 360 GPa\cite{zha_synchrotron_2012}, and it always stays in the insulating molecular phase\cite{goncharov_spectroscopic_2001,loubeyre_optical_2002,zha_synchrotron_2012}.
At high temperatures instead, experimental evidences for an insulator-to-metal transition (IMT) have been reported\cite{weir_metallization_1996} in the range of 100-200 GPa and 2000-3000 K (see Fig.~\ref{f:phase}), altough a clear understanding of the IMT underlying mechanism is still missing, namely if
metallization  occurs via bandgap closure within the molecular phase\cite{nellis_minimum_1999}, or if it is the result of the structural phase transition from a molecular insulating fluid to the conducting atomic one.

Performing experiments under these extreme conditions is really hard, and only few attempts, at high temperatures, have been made\cite{weir_metallization_1996,fortov_phase_2007,dzyabura_evidence_2013}. 
On the other hand, dense liquid hydrogen has been extensively studied within Density Functional Theory (DFT) by molecular dynamics (MD) simulations.
Most of these studies agree on the existence of a first-order transition in the liquid phase between the molecular and the atomic fluid\cite{scandolo_liquidliquid_2003,bonev2004quantum,tamblyn_structure_2010,Morales:2010p28600,morales_nuclear_2013}.
However, its  position in the phase diagram strongly depends on the particular approximation used for the exchange-correlation functional, the difference being of more than one hundred GPa's\cite{Morales:2010p28600,morales_nuclear_2013}.
Moreover a recent systematic study \cite{azadi_fate_2013} clearly demonstrates the inadequacy of standard  approximated DFT functionals
 calculations for solid molecular hydrogen, strongly limiting the DFT predictive power on this subject.

Quantum Monte Carlo (QMC) methods are  among the most accurate and efficient methods for treating many body quantum systems in the thermodynamic limit\cite{foulkes_quantum_2001}. They do not rely on  uncontrolled approximations for the electronic correlation as they allow, in principle, the possibility to reach any desired accuracy, by simply considering correlated wave functions of increasing complexity.
Very recently, and for the first time, in Ref.~\onlinecite{mazzola_unexpectedly_2014} large scale QMC-MD simulations have been reported 
on liquid hydrogen, namely employing long and equilibrated MD simulations, using a system size of 256 protons, comparable with DFT standard ones. 
It has been found that the transition from a molecular fluid towards a fully dissociated one appears at higher pressures with respect to previous DFT and QMC calculations\cite{Morales:2010p28600}.

In this Letter we report QMC-MD simulations based on an efficient generalized second-order Langevin dynamics\cite{krajewski_linear_2006,kuhne_efficient_2007,attaccalite_stable_2008,yeluo_draft_2014} which employ QMC forces, evaluated at the variational Monte Carlo (VMC) level of accuracy Refs~\cite{attaccalite_stable_2008,mazzola_unexpectedly_2014,mazzola_finite-temperature_2012,yeluo_draft_2014}.
Computational details and systematic tests on the accuracy of the trial wavefunction are reported in the Supplementary Material\cite{supmat}.
At variance of Ref.~\onlinecite{mazzola_unexpectedly_2014} we focus on smaller pressures (and with much higher resolution), where an experimental evidence for the IMT has been reported\cite{weir_metallization_1996}.
We indeed find a new transition between the fully molecular (MM) liquid and a mixed molecular-atomic (MA) liquid phase, which appears to be first-order at around 600 K and  440 GPa and is a crossover at about 2300 K and 130 GPa.
The low temperature first order transition line ends in a critical point around 1100-1200 K and $\sim$ 400 GPa.
The MA  phase  eventually dissociates into a fully atomic (AA) liquid  via the first order transition at higher pressures traced in Ref.~\onlinecite{mazzola_unexpectedly_2014}.

 We perform MD simulations in the NVT ensamble exploring a wide range of pressures ($P$) for each isotherm.  
Simulations are long enough well beyond the thermalization time\cite{mazzola_unexpectedly_2014}. We use a supercell containing 256 hydrogen (H) atoms at the $\Gamma$ point, in order to be as close as possible to the thermodynamical limit\cite{tamblyn_structure_2010}. We employ a Jastrow-Slater VMC trial wavefunction as in Ref.~\cite{mazzola_unexpectedly_2014}. The Jastrow part takes into account correlations up to the three-body terms\cite{marchi_resonating_2009,zen_molecular_2013} and the Slater determinant is built with molecular orbitals $\psi_i$ with $i=1,\cdots\,N/2$ (where $N$ is the number of electrons), in turn expanded on an atomic orbitals (AO) basis set.
We use a $2s$ gaussian orbitals/atom basis set which is able to describe almost exactly the single molecule\cite{mazzola_unexpectedly_2014}, but implies a sizable error in the energy for bulk systems.
In order to  study the role of a more accurate ansatz in the liquid phases, we have also adopted a  larger 
basis $3s1p$ which is not far from the
fully converged result with a $6s5p1d$ basis set
 and improves
by about a factor of three the total energy (see Supplementary Materials\cite{supmat}), compared with the smaller 2s-basis set.
The very remarkable outcome that we have obtained in this work is that, even with a small basis set,
the results concerning the radial distribution function and the location of the transition
pressure do not change much (at most by 20 GPa). 

\begin{figure}[htp]
\includegraphics[width=1.0\columnwidth]{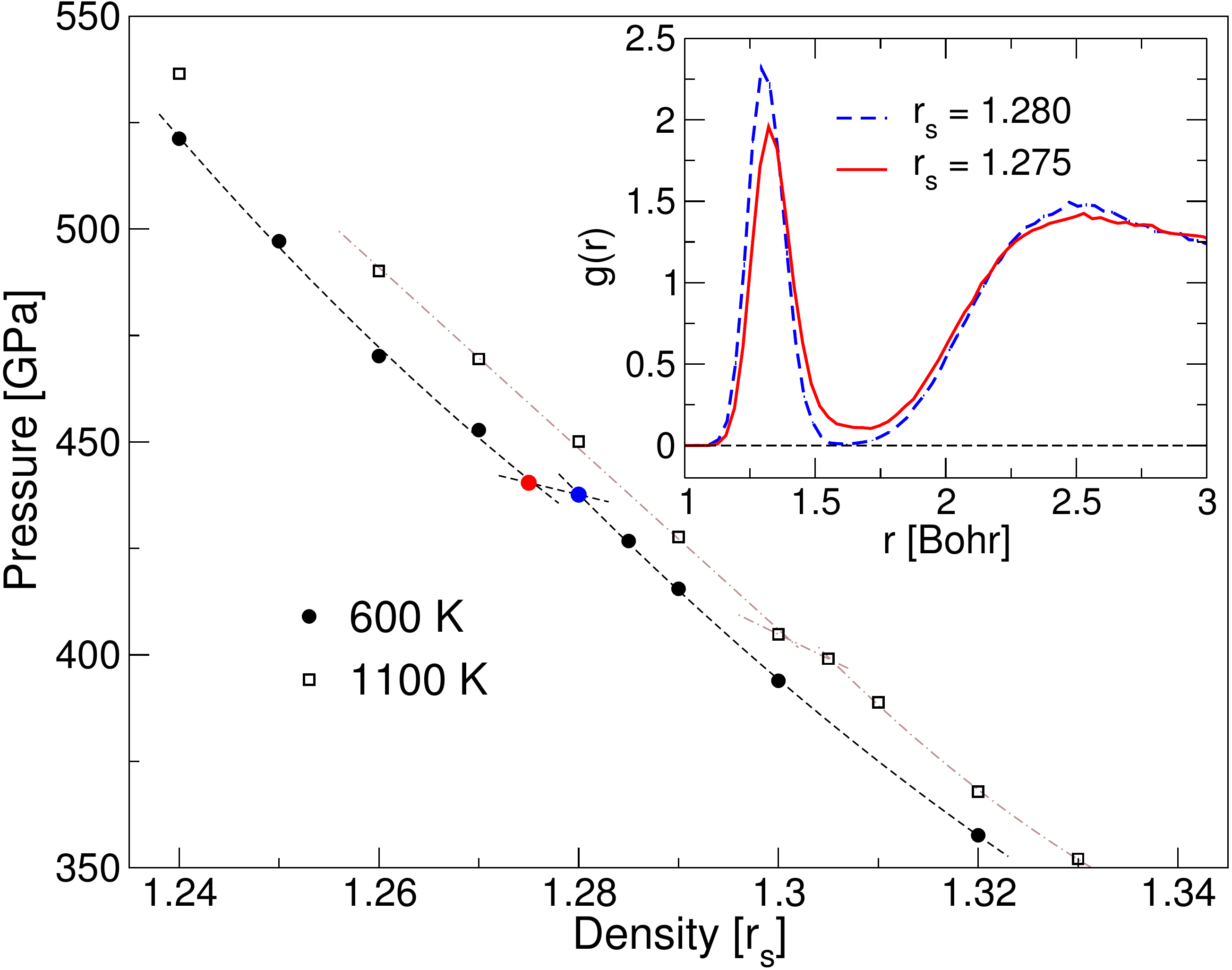}
\caption{ (color online). $P$ vs $\rho$ equations of state for 600 and 1100 K. A clear plateau is visible around $r_s=1.275 - 1.28$ indicating the first order transition at 600 K (full circles), while this plateau is 
less evident at 1100 K (empty squares) and $r_s=1.30$. Error bars are smaller than the symbol's size. Inset: $g(r)$'s  for two densities closest to  the phase boundaries (red and blue points) at 600 K.}
\label{f:eos}
\end{figure}

 In this work we extend the range and mesh of densities explored in Ref.~\onlinecite{mazzola_unexpectedly_2014} and we find a new first-order transition along the 600 K isotherms at $\sim$ 440 GPa, within the liquid phase characterized by the presence of a strong molecular peak in the radial pair distribution function $g(r)$. This transition between the MM and the MA fluids is signalled by a discontinuity in the $P$ vs $\rho$ equation of state (EOS) and also by a qualitative change in the g(r) profile (see Fig.~\ref{f:eos}).
Indeed, starting from 440 GPa, the first minimum of the $g(r)$ is no longer  exactly zero and becomes finite. 
Fluids at densities smaller than $r_s$=1.28\cite{wignerseitz}, show a clear separation in the range of the accessible H-H distances, while all the possible interatomic distances (larger than the molecular equilibrium distance) become allowed when the system is compressed at $r_s<$1.27.
As the density is increased, the region corresponding to the first minimum becomes more and more populated while the height of the first maximum decreases.

\begin{figure}[h!]
\includegraphics[width=\columnwidth]{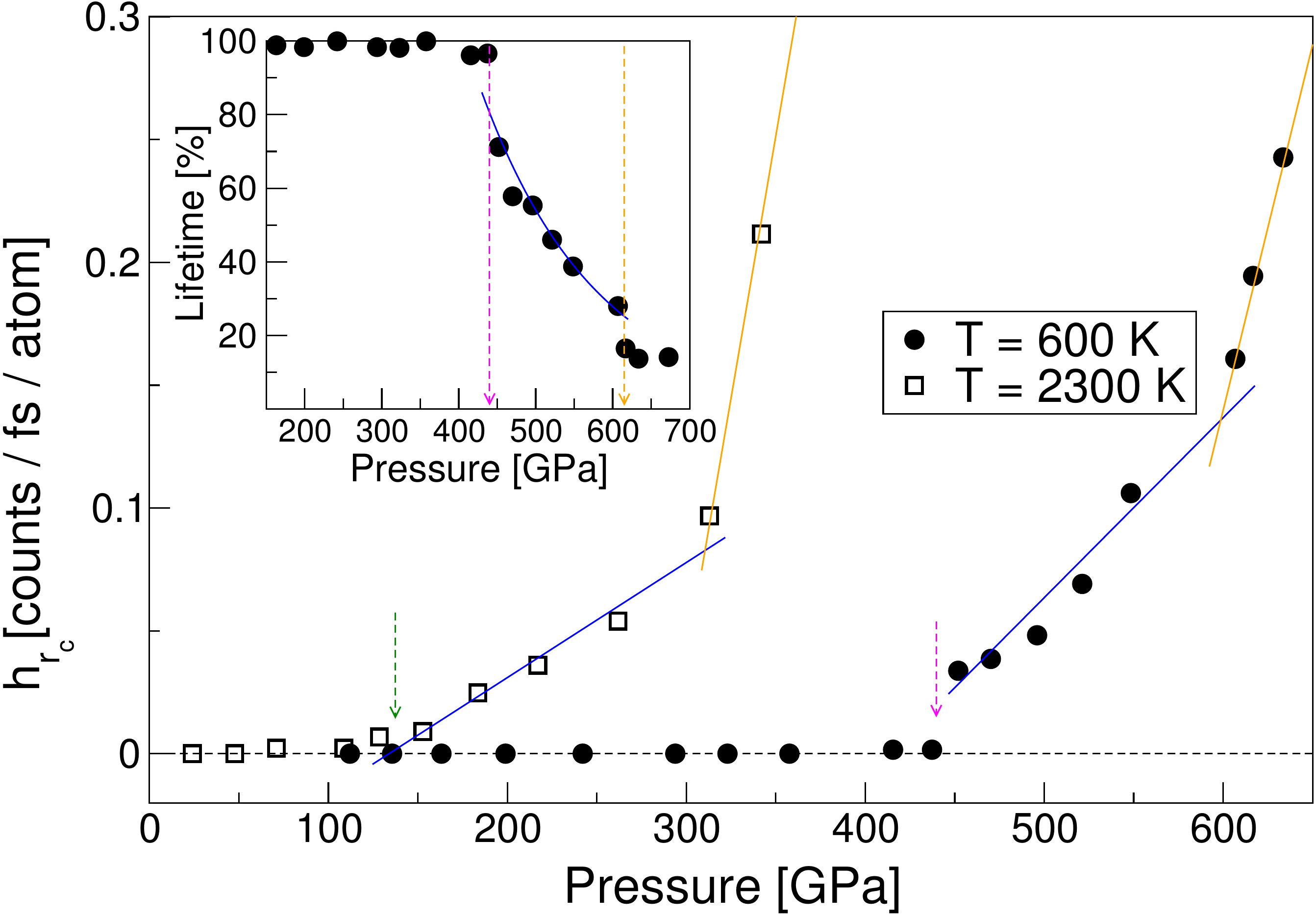}
\caption{ (color online). Recombination rate $h_{r_c=1.6}$ as a function of the pressure and for two isotherms, T=2300 K (600 K) with empty squares (full circles) points.
Blue and orange lines are guides to the eye to distinguish the two phases, namely the  blue 
(orange) line indicates the partially (fully) dissociated liquid. Dashed arrows indicate the largest pressures for which $h_{r_c=1.6}\approx 0$.
 In the inset, the average molecular lifetime, with cutoff $r_c=$ 1.6 Bohr (see text),
 is shown as a function of the pressure at 600 K.
The lifetime is measured in $\%$ with respect to the selected time of observation (400 fs).
}\label{fig:tl}
\end{figure}

The molecular lifetime is a dynamical observable which also displays a clear jump at this pressure.
Altough our Langevin dynamics does not, in principle, allows us to compute dynamical quantities we can still infer interesting properties about the nature of the liquids observed. In Fig.~\ref{fig:tl} (inset) we compute the  lifetime corresponding to bonds shorter than a cutoff $r_c$\cite{cutoff}. 
The qualitative distinction between the three fluids is again evident. The liquid with mixed character, made of short-lived molecules, appears between 440 and 620 GPa.
Let us compute the number $n_{r_c}$ of distinguishable particles which establish a nearest neighbor link (of distance smaller than $r_c$)  to a given atom $i$ during a given simulation time $\Delta t$.
Namely, for each atom $i$ we count how many different hydrogens $j$ form a paired state with $i$ during the simulation. 
In a pure molecular liquid this number should be exactly $n_{r_c}=1$, as all the molecules remain stable along the simulation and there is no proton exchange between the molecules or unpaired atoms in the liquid.
Therefore,  the quantity $h_{r_c} = n_{r_c} - 1$ 
counts how many new $i-j$ bonds\cite{cutoff} are formed during a given time. 
In Fig.~\ref{fig:tl} we observe that the rate becomes finite at 600 K exactly at the first order transition pressure of 440 GPa and diverges near 600 GPa, namely at the complete dissociation\cite{mazzola_unexpectedly_2014}.
A typical recombination rate value for this liquid is of the order of $\sim 0.1$ fs$^{-1}$, which means that the molecular lifetimes are  $\approx$10$^{-14}$ s, i.e, of the order of few molecular oscillations.
The dissociation process appears smoother at 2300 K, nevertheless the recombination becomes noticeably nonzero around $\sim$ 130 GPa,
namely, near the experimental observation of a IMT by Weir et. al.\cite{weir_metallization_1996}.

Within the present QMC calculation we can not directly calculate  the conductivity  as it is common in DFT, since  we do not have direct access to the excited states. Nevertheless, we can infer electronic properties by a careful analysis of the ground state correlation functions\cite{PhysRev.133.A171} within the Born-Oppenheimer approximation.
In particular, since our Jastrow is short range, the long distance behavior of the density matrix can be determined by the uncorrelated density matrix\cite{capello_unconventional_2006} defined as
\begin{equation}
 \Lambda(\br,\br') = \sum_{i=1}^{N/2} \psi_i(\br) \psi_i(\br')
\end{equation}
where $N$ is the number of electrons and $\psi_i$ are the molecular orbitals that define the determinantal part of our wavefunction.
The density matrix decay can be seen by looking at the $r \rightarrow \infty$ behavior of the function $\lambda(r) \equiv \Lambda(0,r)$, where $r = | \br |$. 
The rate of the decay is related
to the size of the energy gap in the band-structure: a larger gap corresponds to a faster decay. In contrast, for a metal
at zero temperature, the decay is much slower, namely power law.
 In a finite size system it is clearly very difficult  to distinguish between different asymptotic behaviors. Nevertheless a qualitative change of the $\lambda(r)$ shape as a function of the density can be observed (see Supplementary Material\cite{supmat}).
\begin{figure}[htp]
\includegraphics[width=1.0\columnwidth]{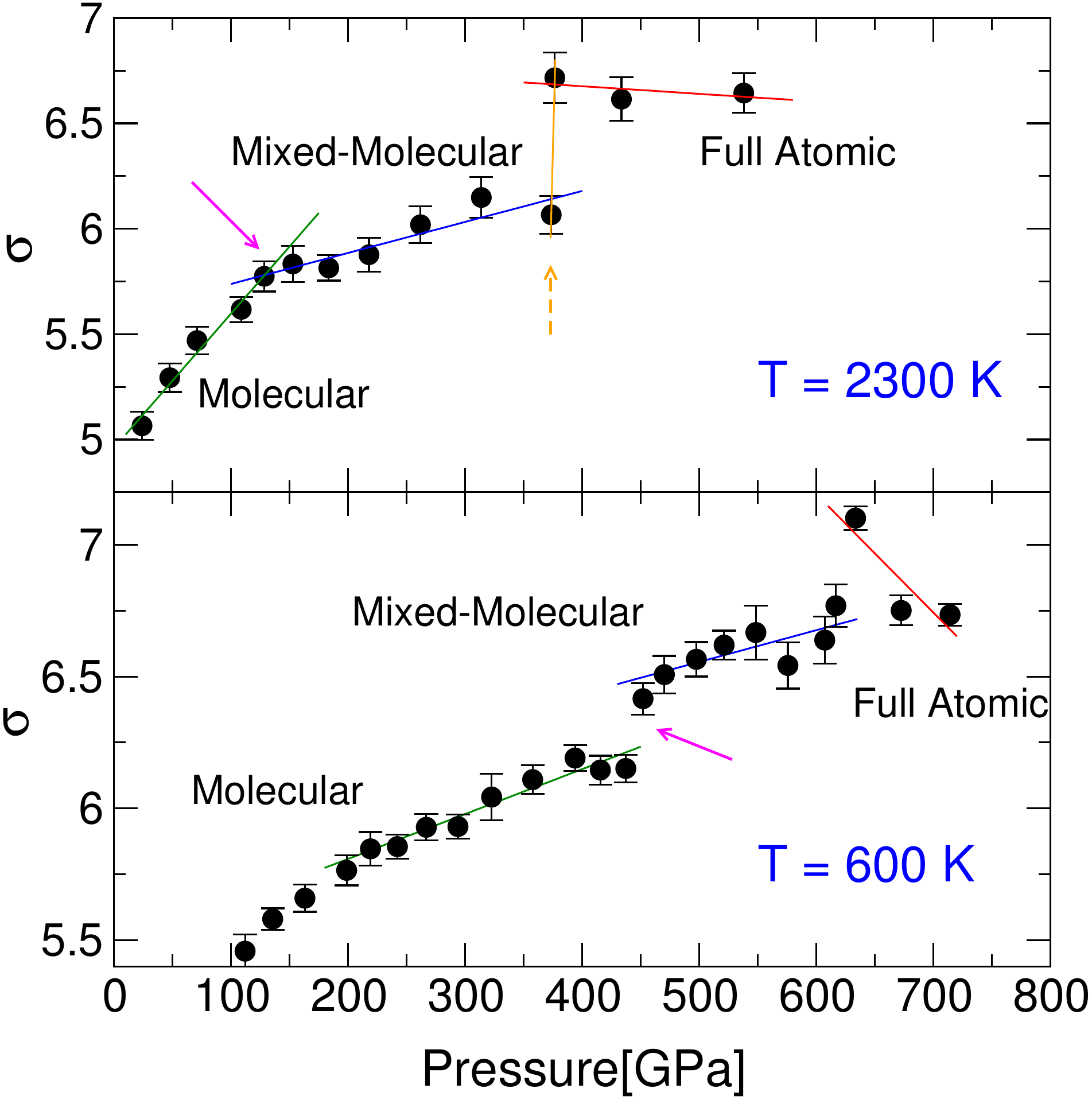}
\caption{ (color online). The density matrix strength  $\sigma$ (infinite for a metal, see text) 
as a function of the pressure for two different isotherms. 
\emph{Top panel:} T=2300 K isotherm. We observe a change in slope (pointed by the pink solid arrow) of $\sigma$ at $\sim$ 130 GPa, before a larger  discontinuity which occurs at 375 GPa, namely at the same pressure of the first order transition  (dashed orange arrow) reported in Ref.~\cite{mazzola_unexpectedly_2014}.
\emph{Top panel:} T=600 K isotherm. This time the discontinuity of $\sigma$ before the complete atomization  is much more clear at $\sim$ 440 GPa (solid pink arrow). At the same pressure a discontinuity in the EOS is shown in Fig.~\ref{f:eos}.
   }
\label{f3:dm}
\end{figure}

To locate the IMT we compute for each density the following quantity
\begin{equation}
 \sigma := \frac1V  \int d{\bm r}' \int d{\bm r} ~| \Lambda(\br', \br) |
\end{equation}
Due to the slow  $\Lambda$ decay in the metal ($ | \Lambda(\br', \br) | \sim |\br^\prime -\br|^{-2} |$), we expect a divergence of this quantity in the metallic phase. 
Since the system is finite, a large jump is hard to observe, but still, some changes are visible passing from an insulating to a metallic phase.
The results are plotted in Fig.~\ref{f3:dm}.
As expected the $ \sigma$ values are larger at high pressures, at which the system should be metallic.
At 2300 K we note a change in the slope  at $\sim$ 130 GPa, before the real first order transition at 375 GPa eventually occurs. 
This value, although its precise connection with the conductivity is still missing, is surprisingly in agreement with the experimental minimum conductivity value at 140 GPa and $\sim$ 2500 K reported in Ref.~\onlinecite{weir_metallization_1996}.
Moreover this is also in agreement that the IM crossover occurs within the molecular phase, as claimed by the authors.
At the lower temperature of 600 K the situation changes, as we observe a more clear jump of the $\sigma$ at $\sim$ 440 GPa, namely 180 GPa's before the transition in which the complete atomization is reached. 
If we identify this point with the IMT, then the metallization at 600 K would occur at the transition between the MM and the MA  liquid, since at this pressure we also observe  the discontinuity in the EOS (see Fig.~\ref{f:eos}) previously discussed.

Summing up, we  
have found, by an advanced ab-initio technique based on QMC, 
a new intermediate phase between the fully atomic and the fully 
molecular liquid phases,  which are commonly assumed in the high-pressure hydrogen phase diagram.
This phase is qualitatively different from 
a  fluid (MM) in which  the molecules - where  the electrons are localized- are somewhat defined (the first maximum of the $g(r)$ being clearly separated by the bulk), and is instead characterized by   a fluid (MA) that favors the electron hopping between the elongated pairs, implying that the system may become metallic at these densities.
This transition appears to be first order at 600 K and is smoothed with increasing temperature. The critical end point is located just above 1100 K, because the first order transition is very weak 
 in this case (see Fig.~\ref{f:eos}).

In Fig.\ref{f:phase} we update the high pressure phase diagram of Ref.~\onlinecite{mazzola_unexpectedly_2014} with the data concerning the IMT and the region  of stability of this MA liquid.
Notice that our proposed IMT line is surprisingly in agreement with the one  obtained by DFT simulations with HSE functional  reported in
 Ref.~\onlinecite{morales_nuclear_2013}.
However, this agreement  could be only accidental because in Ref.~\onlinecite{morales_nuclear_2013} it is not specified if IMT and the complete atomization occurs simoultaneously in the simulations.
 In our QMC simulations instead,  we can distinguish between three, qualitatively different, types of 
 liquids.

In the MA liquid, the coordination shells are not well separated but the persistence of a broad static molecular peak is evident.
This new phase is exceptionally stable, as  a nonzero molecular fraction has been observed\cite{mazzola_unexpectedly_2014} up to 375 GPa at 2300 K and 620 GPa at 600 K.
This fluid is also characterized by the fact that the molecules are short-lived compared with  the fully molecular one. Dissociation and recombination events occur on a timescale comparable to few molecular oscillation periods.
By looking at decay of the density matrix, computed with very accurate single particle orbitals obtained within the VMC calculation, we propose that this liquid should have metallic behavior.
Thus the metallization at finite temperature occurs within the (partially) molecular phase.
At 2300 K the metallization takes place at a pressure $\sim$130 GPa, a value which is very well in agreement with the experiment, while at 600 K the IMT is predicted to be at $\sim$440 GPa.
Finally, the fluid completely dissociates into a fully atomic liquid (AA) phase.

All these outcomes have been obtained with the classical nuclei approximation, therefore the quantum proton corrections are expected to shift the two transitions towards smaller pressures, especially at lower temperatures.
Therefore, altough the quest for a pure atomic hydrogen compound  seem to be out of reach for the present experimental possibilities, achieving the low temperature metallization of fluid molecular hydrogen appears more realistic in the next future.

\acknowledgments
Computational resources were provided by K computer at RIKEN Advanced Institute for 
Computational Science (AICS).
We  thank M. Holzmann and S. Yunoki for useful discussions. 
We also thank C. Pierleoni for providing protonic configurations and energies obtained with the CEIMC technique.
This work is partially supported by MIUR-PRIN-2010.



\begin{thebibliography}{10}
\bibitem{wignerseitz}
$r_s$ is the Wigner-Seitz radius defined by $V/N= 4/3 \pi (r_s a_0)^3$ where
  $V$ is the volume, $N$ the number of ions, and $a_0$ is the Bohr radius.
\bibitem{cutoff}
we adopt a cutoff of $r = 1.6$ Bohr (the results do not depend qualitatively by
  the particular choice of this cutoff).
\end{thebibliography}
\end{document}